\documentclass[11pt,a4paper]{article}
\usepackage[top=1.0in, bottom=1.0in, left=1.0in, right=1.0in]{geometry}
\usepackage{authblk}
\usepackage{amssymb, amsmath,framed}
\usepackage{cancel}
\usepackage{amsfonts}
\usepackage[T1]{fontenc}
\usepackage{bm}
\usepackage[round]{natbib}
\usepackage{graphics,graphicx}
\usepackage{aastex_hack}
\usepackage{comment}
\usepackage{threeparttable}
\usepackage[compact,tiny]{titlesec}
\usepackage{esvect}
\usepackage{floatrow}
\usepackage{tabularx}
\usepackage{hyperref}
\usepackage{color}
\usepackage{caption}
\DeclareSymbolFont{operators}{OT1}{cmr}{m}{n}
\DeclareSymbolFont{letters}{OML}{cmm}{m}{it}
\DeclareSymbolFont{symbols}{OMS}{cmsy}{m}{n}
\DeclareSymbolFont{largesymbols}{OMX}{cmex}{m}{n}

\usepackage{hyperref}
\hypersetup{
    bookmarks=true,         
    unicode=false,          
    pdftoolbar=true,        
    pdfmenubar=true,        
    pdffitwindow=false,     
    pdfstartview={FitH},    
    pdfsubject={Plasma fluid theory},   
    pdfproducer={GNU Make}, 
    pdfkeywords={keyword1} {key2} {key3}, 
    pdfnewwindow=true,      
    colorlinks=true,        
    linkcolor=blue,      
    citecolor=blue,         
    filecolor=blue,      
    urlcolor=blue           
}
\usepackage{array}

\def\ga{\gamma}

\makeatother

\begin{document}
\title{Atmospheric escape from the TRAPPIST-1 planets and implications for habitability}
\author{Chuanfei Dong\thanks{To whom correspondence should be addressed. E-mail: \texttt{dcfy@princeton.edu}}}
\affil{Department of Astrophysical Sciences, Princeton University, Princeton, NJ 08544, USA}
\affil{Princeton Center for Heliophysics, Princeton Plasma Physics Laboratory, Princeton University, Princeton, NJ 08544, USA}
\author{Meng Jin}
\affil{Lockheed Martin Solar and Astrophysics Lab, Palo Alto, CA 94304, USA}
\author{Manasvi Lingam}
\affil{Harvard-Smithsonian Center for Astrophysics, Cambridge, MA 02138, USA}
\affil{John A. Paulson School of Engineering and Applied Sciences, Harvard University, Cambridge, MA 02138, USA}
\author{Vladimir S. Airapetian}
\affil{NASA Goddard Space Flight Center, Greenbelt, MD, USA}
\author{Yingjuan Ma}
\affil{Institute of Geophysics and Planetary Physics, University of California, Los Angeles, CA 90095}
\author{Bart van der Holst}
\affil{Center for Space Environment Modeling, University of Michigan, Ann Arbor, MI 48109, USA}

\date{}

\maketitle

\begin{abstract}
The presence of an atmosphere over sufficiently long timescales is widely perceived as one of the most prominent criteria associated with planetary surface habitability. We address the crucial question as to whether the seven Earth-sized planets transiting the recently discovered ultracool dwarf star TRAPPIST-1 are capable of retaining their atmospheres. To this effect, we carry out numerical simulations to characterize the stellar wind of TRAPPIST-1 and the atmospheric ion escape rates for all the seven planets. We also estimate the escape rates analytically and demonstrate that they are in good agreement with the numerical results. We conclude that the outer planets of the TRAPPIST-1 system are capable of retaining their atmospheres over billion-year timescales. The consequences arising from our results are also explored in the context of abiogenesis, biodiversity, and searches for future exoplanets. In light of the many unknowns and assumptions involved, we recommend that these conclusions must be interpreted with due caution.
\end{abstract}


With the number of detected exoplanets now exceeding 3600 \citep{WF15}, exoplanetary research has witnessed many remarkable advances recently. One of the most important areas in this field is the hunt for Earth-sized terrestrial planets residing in the habitable zone (HZ) of their host stars - the HZ represents the region within which a planet can support liquid water on its surface \citep{Kop13}; a probabilistic version of the HZ, encompassing a wide range of planetary and stellar parameters, has also been formulated \citep{Zs15}. The importance of this endeavour stems from the fact that such planets can be subjected to further scrutiny to potentially resolve the question as to whether they may actually harbour life \citep{Cock16}.

Most of the recent attention has focused on exoplanets in the HZ of M-dwarfs, i.e. low-mass stars that are much longer lived than the Sun, for the following reasons. Firstly, M-dwarfs are the most common type of stars within the Milky Way \citep{Chab03}, implying that $\sim 10^{10}$ Earth-sized planets in the HZ of M-dwarfs may exist in our Galaxy \citep{DC15}. Second, owing to the HZ being much closer to such stars, it is much easier to detect exoplanets and characterize their atmospheres, if they do exist \citep{SBJ16}. Lastly, this field has witnessed two remarkable advances within the last year: the discovery of Proxima b \citep{AE16} and the seven Earth-sized planets transiting the ultracool dwarf TRAPPIST-1 \citep{Gill16,Gill17}. The significance of the former stems from the fact that it orbits the star closest to the Solar system, and the latter is important because there exist as many as three planets in the HZ with the possibility of life being seeded by panspermia \citep{LL17}.

In light of these discoveries, the question of whether terrestrial exoplanets in the HZ of M-dwarfs are habitable is an important one \citep{SBJ16}. Amongst the many criteria identified for a planet to be habitable, the existence of an atmosphere has been posited as being crucial for surficial life-as-we-know-it \citep{Lam09,Cock16}. It is therefore evident that the study of atmospheric losses from exoplanets constitutes a crucial line of enquiry. Empirical and theoretical evidence from our own Solar system suggests that the erosion of the atmosphere by the stellar wind plays a crucial role, especially for Earth-sized planets where such losses constitute the dominant mechanism \citep{Lammer13,brain16} and the same could also be true for exoplanets around M-dwarfs \citep{DLMC,ZC17}. Recent studies of atmospheric ion escape rates from Proxima b (and other M-dwarf exoplanets) also appear to indicate that the resulting ion losses are significant because of the extreme space weather conditions involved \citep{GDC16}, potentially resulting in the atmosphere being depleted over a span ranging from tens to hundreds of millions of years \citep{DLMC,WW,Aira17,Garcia-Sage17}.

Hence, in this paper we shall focus primarily on the atmospheric ion escape rates of the seven TRAPPIST-1 planets by adapting a sophisticated multi-species magnetohydrodynamics (MHD) model which self-consistently includes ionospheric chemistry and physics, and electromagnetic forces. In this work, we do not tackle the wide range of hydrodynamic escape mechanics that have been explored for terrestrial planets \citep{OA16,ZC17} on account of the above reasons.

\begin{figure*}
\begin{center}
\includegraphics[scale=0.51]{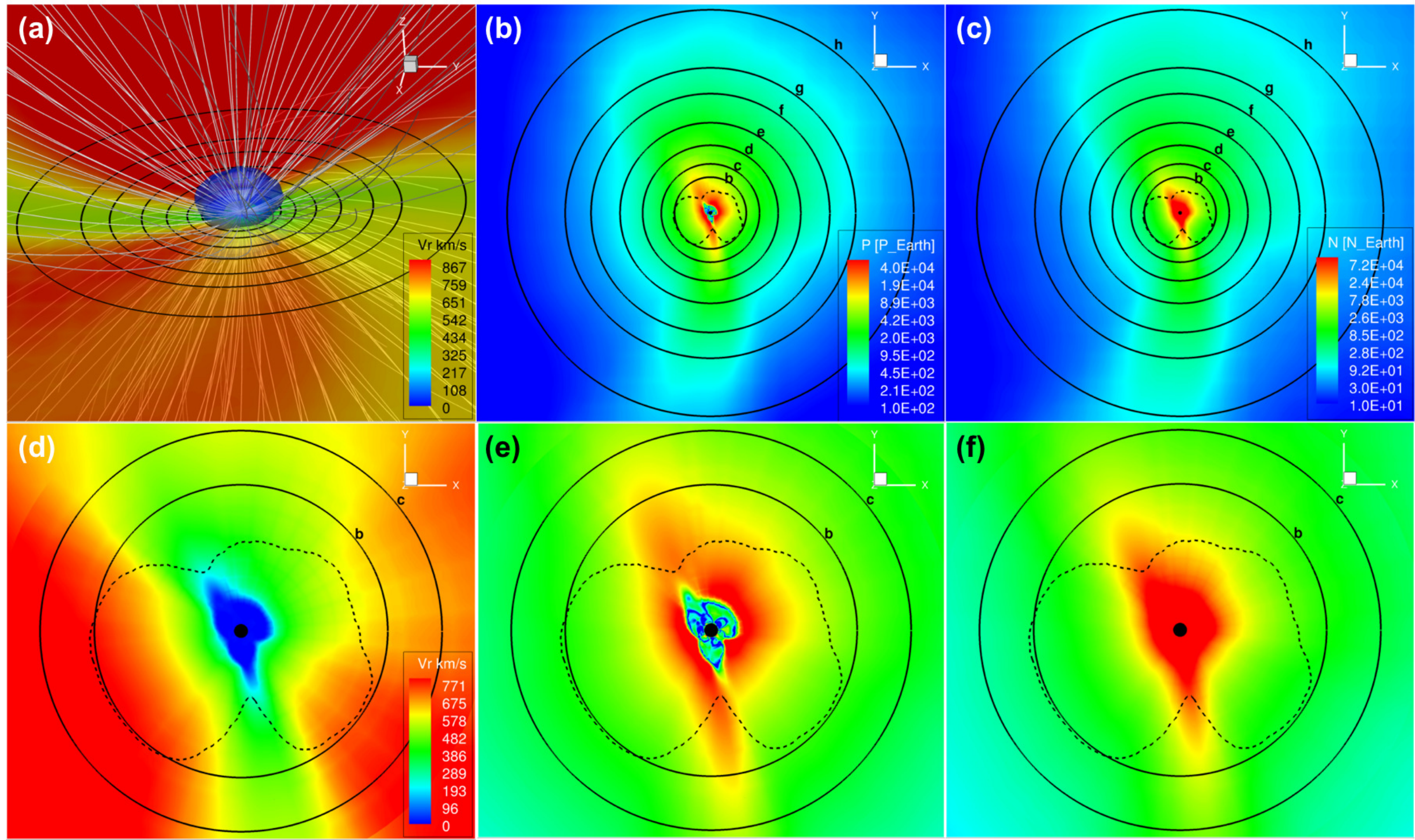}
\end{center}
\caption{The steady state stellar wind of TRAPPIST-1. (a) The 3D stellar wind configuration with selected magnetic field lines. The background contour shows stellar wind speed at the equatorial plane (z=0). The blue isosurface represents the critical surface beyond which the stellar wind becomes super-magnetosonic. The black solid lines represent the orbits of seven planets TRAPPIST-1b to TRAPPIST-1h. (b) The equatorial plane z=0 showing the stellar wind dynamic pressure normalized by the solar wind dynamic pressure at 1 AU. The dashed line shows the critical surface location. (c) The equatorial plane z=0 showing the stellar wind density normalized by the solar wind density at 1 AU. (d)-(f) A zoom-in view of the equatorial plane at z=0 near TRAPPIST-1 depicting the stellar wind velocity, normalized dynamic pressure, and density, respectively. Note that the color bar for (e) is the same as (b), and that of (f) is identical to (c).}
\label{fig:stellarwind}
\end{figure*}

\section*{The stellar wind of TRAPPIST-1}
In order to commence our analysis of stellar wind-induced atmospheric loss, the stellar wind parameters of TRAPPIST-1 are required. Since the space weather conditions at the TRAPPIST-1 planets are presently unknown from observations, we must rely upon simulating the stellar wind of TRAPPIST-1. The latter is implemented by means of the Alfv\'{e}n Wave Solar Model (AWSoM), a data-driven global MHD model that was originally developed for simulating the solar corona and solar wind \citep{SVO13,bart14}. The AWSoM has been proven to be successful in reproducing high-fidelity solar corona conditions \citep{bart14, oran13}, and can readily be adapted to self-consistently model stellar wind profiles for a wide range of stars \citep{vidotto13,cohen14,GDC16}. To adapt the AWSoM for modelling the TRAPPIST-1 stellar wind, we utilize the rotational period, radius, and mass of the star based on the latest estimates \citep{Luger17} and a mean magnetic field typical of similar late M-dwarfs \citep{morin10}. Further details concerning our approach can be found in Appendix \ref{AppA}.

The steady state stellar wind solution is illustrated in Fig. \ref{fig:stellarwind}. When compared to the normal solar wind solution \citep{jin12}, the stellar wind of TRAPPIST-1 is much faster ($\sim 3$ times) when evaluated at the same stellar distance. The critical surface is defined as the region where $v_{s}=v_{f}$, with $v_{s}$ and $v_{f}$ representing the stellar wind and fast magnetosonic speeds respectively. The surface occurs at distances of $\sim 30$ $R_{*}$ and $\sim 20$ $R_{*}$ for the fast and slow stellar wind; the latter duo originate at different regions on the star and have different speeds and densities \citep{bart14}. Hence, this leads to a very unique feature of the TRAPPIST-1 system, namely that part of the orbit of TRAPPIST-1b (the closest planet) lies \emph{within} the critical surface, while all the other planets are embedded in the super-magnetosonic stellar wind. 

This striking scenario does not exist within our own solar system primarily because of the proximity of TRAPPIST-1b to its host star (in conjunction with a strongly magnetized stellar wind). As TRAPPIST-1b orbits within the critical surface, the planet could magnetically interact with its host star directly. In turn, the star-planet interaction could perhaps (i) regulate the rotational rate \citep{Pont09}, (ii) modify the properties of a local dynamo \citep{CSM00}, and (iii) even give rise to a dynamo mechanism \citep{CeHo14}. In this context, we observe that variations in the magnetic field occur during the stellar cycle caused by the dynamo process. Thus, the distance of the critical surface is expected to also vary concomitantly, implying that TRAPPIST-1b could be subject to frequent transitions between sub-magnetosonic and super-magnetosonic stellar wind conditions along the lines of Proxima b \citep{GDC16}.

Another distinguishing feature of the stellar wind from TRAPPIST-1 is its higher density. When combined with the higher wind speed, all of the planets are subjected to a much larger dynamic pressure compared to that experienced by the Earth. At the orbit of TRAPPIST-1b, the dynamic wind pressure is about $10^3$-$10^4$ times greater than the solar wind dynamic pressure at Earth. Even when we consider the furthermost planet, TRAPPIST-1h, the dynamic pressure is about $100$-$300$ times larger than the near-Earth environment. The existence of such an extreme wind pressure has already been documented for Proxima b \citep{GDC16}, and its effects on the evolution of the planet's magnetosphere have also been thoroughly investigated \citep{Aira17,DLMC}. The ramifications of these extreme space weather conditions on the atmospheric ion escape rates of the TRAPPIST-1 planets are explored in the subsequent section. 

We note that the mass-loss rate from TRAPPIST-1 is $\sim 2.6 \times 10^{11}$ g/sec, which is about $10\%$ of the solar mass-loss rate. Although the density and velocity of the stellar wind are higher for TRAPPIST-1, the smaller size of the host star is responsible for yielding a value lower than that of the active young Sun \citep{AU16}. The mass-loss rate $\left(\sim 0.1 \dot{M}_\odot\right)$ obtained for TRAPPIST-1 is broadly consistent with the upper bound of $\sim 0.2 \dot{M}_\odot$ for the slightly larger star, Proxima Centauri \citep{WLMZ}.

Lastly, the stellar wind parameters provided in this paper are useful in determining the radio auroral emission from the TRAPPIST-1 planets, which can be used to constrain their magnetic fields \citep{Gri15}. In Appendix \ref{AppB}, we show that the radio emission could potentially peak at $\mathcal{O}(0.1)$ MHz, and result in a radio flux density of $\mathcal{O}(0.1)$ mJy; the latter could be enhanced by $2$-$3$ orders of magnitude during a Coronal Mass Ejection (CME) event. 

\section*{Ion escape rates for the TRAPPIST-1 planets}
To simulate the ion escape rates for the seven planets of the TRAPPIST-1 system, we employ the sophisticated 3-D Block Adaptive Tree Solar-wind Roe Up-wind Scheme (BATS-R-US) multi-species MHD (MS-MHD) model that has been extensively tested and validated in the Solar system for Venus and Mars \citep{Ma04,Toth12,Ma13,Jak15}, and was recently employed to study the atmospheric losses from Proxima b \citep{DLMC}. The reader is referred to these papers and the Appendix for further details concerning the details of the numerical implementation, the model equations, and the physical and chemical processes encoded within the model. Note that the neutral atmosphere is the source of the produced ions through, e.g., photoionization and charge exchange, and that only a small fraction of them will escape into space. Hence, the atmosphere will, in addition to being eroded, also undergo changes in the chemical composition \citep{DLMC}.

\subsection*{The input parameters of the model}

We shall, for the most part, concern ourselves with describing and motivating our choice of the different input parameters required for the BATS-R-US model. Before proceeding further, we wish to caution the readers that many of the relevant planetary and stellar wind parameters of the TRAPPIST-1 system are unknown or poorly constrained. Hence, it is important to recognize that, on account of the many uncertainties involved, the ensuing escape rates may not necessarily be representative of the TRAPPIST-1 system.

The BATS-R-US model relies upon an atmospheric composition akin to Venus and Mars, implying that the TRAPPIST-1 planets are also assumed to possess a similar composition. There are several factors that must be noted in this context. Firstly, as seen from our Solar system, the ion escape rates for Venus, Mars and Earth are similar despite their compositions, sizes and magnetic field strengths being wildly dissimilar \citep{Lammer13,brain16}, thereby indicating that the ion escape rates may be relatively sensitive to stellar wind parameters compared to planetary properties\footnote{The difference in stellar wind parameters at Venus, Earth and Mars is only up to a $\mathcal{O}(1)$ factor.}; this is also partly borne out by the atmospheric ion escape rate calculations for Proxima b \citep{Aira17,DLMC}. In addition, it has been shown recently that the ion escape rates are only weakly dependent on the surface pressure \citep{DLMC}. We observe that the inner planets of the TRAPPIST-1 system could have experienced significant losses of $H_2$ and water over fast timescales \citep{Bel17,Bour17}, leaving behind other atmospheric components. Lastly, a Venus-like atmosphere for the inner planets cannot be ruled out empirically, as noted in \citep{DeW17}.\footnote{In broader terms, gaining a thorough understanding of Venus-type exoplanets is highly relevant, because it allows us to compare and contrast their properties against exo-Earths \citep{Ang17}.}

The next two input parameters to be specified are the surface pressure and the scale height for each of the planets. The former remains unknown at this stage, and we work with the fiducial value of $1$ atm at this stage. We anticipate that the surface pressure does not significantly alter the escape rates, at least for extreme stellar wind conditions, as demonstrated in \citep{DLMC}. The scale height $H_x$ is defined as
\begin{equation}
    H_x = \frac{k T}{m g_x},
\end{equation}
where $g_x$ is the acceleration due to gravity for planet $X$. The latter quantity can be easily computed for all the planets since their masses and radii are known \citep{Gill17,Luger17}.

\begin{table}
\caption{Ion escape rates in sec$^{-1}$.}\label{tableesc}
\centering
\begin{tabular}{lllll}
\hline
& O$^+$ & O$_2^+$ & CO$_{2}^+$ & Total  \\
\hline
\multicolumn{5}{c} {Maximum total pressure} \\
\hline
Trappist-1b & 5.56$\times$10$^{27}$  & 2.09$\times$10$^{26}$   &  1.52$\times$10$^{26}$   &  5.92$\times$10$^{27}$ \\ 
Trappist-1c & 1.54$\times$10$^{27}$  &  1.38$\times$10$^{26}$  &   1.32$\times$10$^{26}$  &  1.81$\times$10$^{27}$ \\ 
Trappist-1d & 1.29$\times$10$^{27}$  &  3.80$\times$10$^{25}$  &   1.14$\times$10$^{25}$  &  1.34$\times$10$^{27}$ \\
Trappist-1e & 7.01$\times$10$^{26}$  &  2.83$\times$10$^{25}$  &   1.10$\times$10$^{25}$  &  7.40$\times$10$^{26}$ \\ 
Trappist-1f & 5.23$\times$10$^{26}$  &  3.37$\times$10$^{25}$  &   1.19$\times$10$^{25}$  &  5.68$\times$10$^{26}$ \\ 
Trappist-1g & 2.17$\times$10$^{26}$  &  2.71$\times$10$^{25}$  &   1.32$\times$10$^{25}$  &  2.58$\times$10$^{26}$ \\ 
Trappist-1h & 1.06$\times$10$^{26}$  &  1.65$\times$10$^{25}$  &   6.98$\times$10$^{24}$  &  1.29$\times$10$^{26}$ \\ 
\hline
\multicolumn{5}{c} {Minimum total pressure} \\ 
\hline
Trappist-1b & 9.33$\times$10$^{26}$  & 4.99$\times$10$^{25}$   &  2.92$\times$10$^{25}$   &  1.01$\times$10$^{27}$ \\ 
Trappist-1c & 4.23$\times$10$^{26}$  &  9.22$\times$10$^{25}$  &   2.76$\times$10$^{25}$  &  5.42$\times$10$^{26}$ \\ 
Trappist-1d & 2.81$\times$10$^{26}$  &  3.07$\times$10$^{25}$  &   1.04$\times$10$^{25}$  &  3.23$\times$10$^{26}$ \\
Trappist-1e & 2.20$\times$10$^{26}$  &  4.19$\times$10$^{25}$  &   1.25$\times$10$^{25}$  &  2.74$\times$10$^{26}$ \\ 
Trappist-1f & 1.88$\times$10$^{26}$  &  4.30$\times$10$^{25}$  &   1.10$\times$10$^{25}$  &  2.42$\times$10$^{26}$ \\ 
Trappist-1g & 9.33$\times$10$^{25}$  &  5.85$\times$10$^{25}$  &   1.38$\times$10$^{25}$  &  1.66$\times$10$^{26}$ \\ 
Trappist-1h & 4.52$\times$10$^{25}$  &  2.69$\times$10$^{25}$  &   4.39$\times$10$^{24}$  &  7.66$\times$10$^{25}$ \\ 
\hline
\end{tabular}
\end{table}

The stellar wind parameters are obtained from the model described in the previous section. We must also prescribe the extreme ultraviolet (EUV) fluxes received at each of these planets, since the EUV flux plays an important role in regulating the extent of photoionization and the resultant stellar heating. This is accomplished by utilizing the values for the TRAPPIST-1 planets computed in \citep{Bel17,Wheat17,Bour17}.

The planetary magnetic field is a potentially important factor in regulating the ion escape rates. We consider the scenario where the planets are unmagnetized because this case yields an upper limit on the allowed escape rates \citep{DLMC}. Hence, if a planet was characterized by ``low'' escape rates in the unmagnetized limit, it would also typically possess low escape rates in the presence of a magnetic field. It must also be borne in mind that the planets orbiting TRAPPIST-1 are likely to be tidally locked, and it has been argued that such planets are likely to possess weak magnetic fields \citep{Khoda07}. If a planet is weakly magnetized, it is likely that the total ion escape rate will be comparable to (but slightly lower than) the unmagnetized case \citep{DLMC}.

\begin{figure*}
\begin{center}
\includegraphics[scale=0.4]{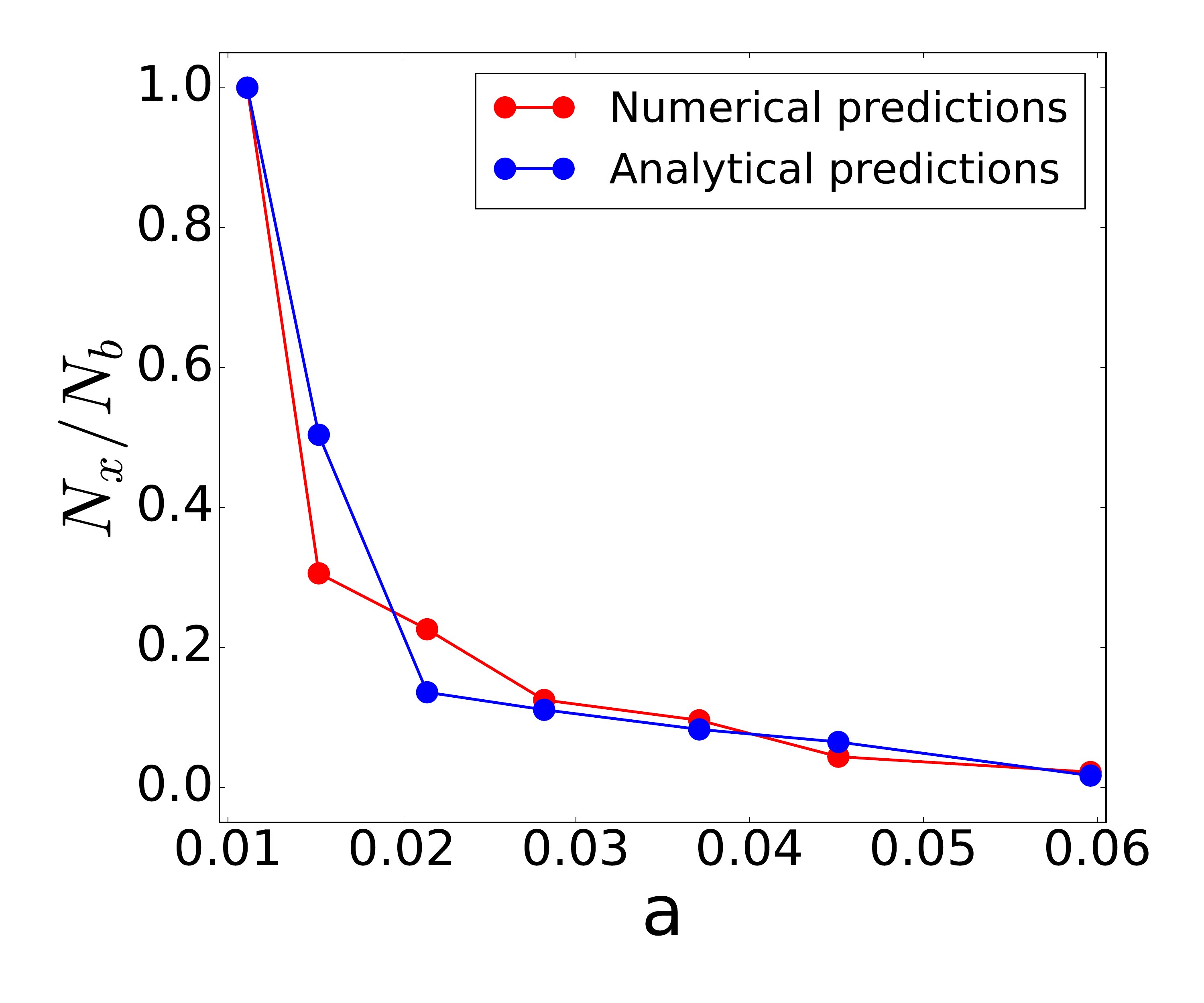}
\end{center}
\caption{The plot of the normalized escape rate as a function of the semi-major axis for the case with maximum total pressure. The seven distinct points represent the seven planets of the TRAPPIST-1 system.}
\label{figescrate}
\end{figure*}

\begin{figure*}
\begin{center}
\includegraphics[scale=0.75]{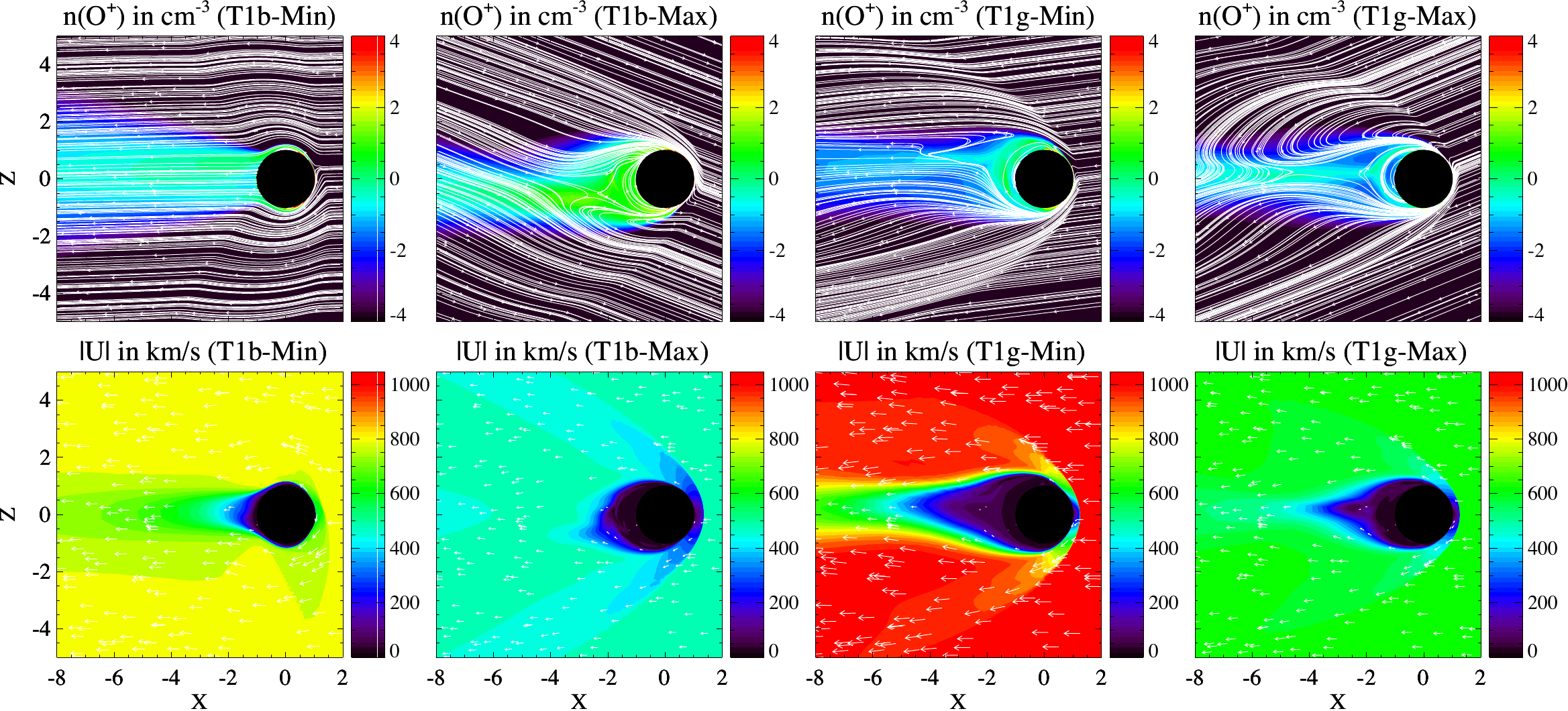}
\end{center}
\caption{The logarithmic scale contour plots of the O$^+$ ion density (first row) with magnetic field lines (in white) and stellar wind speed (second row) with stellar wind velocity vectors (in white) in the meridional plane. The first two columns correspond to TRAPPIST-1b (minimum and maximum P$_{tot}$) while the last two represent TRAPPIST-1g (minimum and maximum P$_{tot}$). The X- and Z-coordinates have been normalized in terms of the corresponding planetary radius.}
\label{figcontplot}
\end{figure*}

\subsection*{Results from the model}
For each planet, we consider two limiting cases. The first corresponds to the scenario with \emph{maximum} dynamic (and total) pressure over one orbit of the planet. The second corresponds to the case with \emph{minimum} total pressure, but with the \emph{maximum} magnetic pressure. The corresponding stellar wind parameters have been provided in the Appendix, and the escape rates in Table \ref{tableesc}. 

For all seven planets, the case with maximum total pressure yields a total atmospheric ion escape rate that is a few times higher than the corresponding case with minimum total pressure. The innermost trio of planets (`b', `c' and `d') have escape rates higher than $10^{27}$ sec$^{-1}$, while the last four have rates lower than this value when the case with maximum total pressure is considered. In comparison, the escape rates for Mars, Venus and Earth are $\sim 10^{24}-10^{25}$ sec$^{-1}$ \citep{Lammer13}, while that of Proxima b is $\sim 2 \times 10^{27}$ \citep{DLMC}. In our subsequent analysis, we shall focus on this case (maximum total pressure) since it leads us towards determining the upper bounds on the escape rates.

Using the mixing-length formalism of \citep{CR91} in conjunction with the definitions of the stellar and planetary mass-loss rates, it can be shown that
\begin{equation} \label{AnMassRat}
N_x \propto \left(\frac{R_x}{a}\right)^2 \dot{M}_\star,
\end{equation}
where $N_x$ is the atmospheric escape rate arising from stellar wind stripping, $\dot{M}_\star$ is the stellar mass-loss rate, $R_x$ and $a$ are the radius and semi-major axis of the planet $X$ respectively \citep{ZSR10,LiLo17}. We have normalized the escape rates for the TRAPPIST-1 planets in terms of the escape rate for TRAPPIST-1b, and compared the numerical and analytical predictions in Fig. \ref{figescrate}. 

An inspection of Fig. \ref{figescrate} reveals that the analytical formula is in good agreement with the numerical simulations, especially for the last four planets (`e' to `h') which are regarded as being potentially capable of retaining atmospheres over Gyr timescales. Even for the inner planets, we find that the analytical results fall within the numerical values by a factor of $\lesssim 2$. Hence, Eq. \ref{AnMassRat} may facilitate a quick estimation of the atmospheric escape rates from unmagnetized planets. However, for both the simulations and the analytic results, it is important to note that the atmospheric escape rates were higher in the past, although quantitative estimates are difficult since the time-dependent stellar mass-loss rates are poorly constrained.

Fig. \ref{figcontplot} depicts how the oxygen ions escape from two planets of the TRAPPIST-1 system (`b' and `g') along with the interplanetary magnetic field lines. From the color contours, it is evident that the ion loss rate at TRAPPIST-1b is larger than TRAPPIST-1g. The solar wind magnetosonic Mach number of TRAPPIST-1b for the case with minimum total pressure (P$_{tot}$) equals $0.94$, therefore there is no shock formed in front of the planet (because of its sub-magnetosonic nature). This is particularly noteworthy since this condition is \emph{not} prevalent in our Solar system for any of the eight planets. On the other hand, such a condition has been observed for moons located inside the planetary global magnetosphere (e.g., Ganymede and Titan). In contrast to the above case, the bow shock is present in all other cases and planets of the TRAPPIST-1 system.

Based on the background colors of the second row, minimum P$_{tot}$ corresponds to the fast stellar wind and maximum P$_{tot}$ signifies the slow stellar wind, which is fully consistent with the classical solar/stellar wind picture. The high P$_{tot}$ manifested in the `slow' stellar wind is mostly a consequence of the high stellar wind density. The second row of Fig. \ref{figcontplot} also indicates that the stellar wind accelerates away from the star; the stellar wind velocity at TRAPPIST-1g is obviously larger than that of TRAPPIST-1b. These conclusions can also be verified through an inspection of Table \ref{tableSW} in the Appendix. 

From Fig. \ref{figcontplot}, we observe that the plasma boundaries are compressed because of the fact that the total pressure is much higher at these planets compared to that experienced by Venus. The escape of ions is primarily driven by the pressure gradient and ${\bf J} \times {\bf B}$ forces in the momentum equation \citep{Ma13}. Lastly, we refer the reader to Fig. \ref{figIonProf} in the Appendix that depicts the ionospheric profiles of TRAPPIST-1g for the cases with minimum and maximum total pressure.

\section*{Discussion and Conclusions}
We have arrived at an upper bound on the ion escape rates by considering the scenario where the planets are unmagnetized and subject to the maximum total (or dynamic) pressure. It was concluded that the innermost planet TRAPPIST-1b has an upper bound of $5.92 \times 10^{27}$ sec$^{-1}$ while the corresponding value for the outermost planet TRAPPIST-1h is $1.29 \times 10^{26}$ sec$^{-1}$. As the planets are approximately Earth-sized and assumed to have a surface pressure of $1$ atm, we can estimate the timescales over which these planets can retain their atmospheres. The values range from $\mathcal{O}\left(10^8\right)$ years for TRAPPIST-1b to $\mathcal{O}\left(10^{10}\right)$ years for TRAPPIST-1h. Moreover, we also see from Fig. \ref{figescrate} that the overall escape rate declines monotonically as one moves outwards, from TRAPPIST-1b to TRAPPIST-1h. Hence, taken collectively, this may suggest that TRAPPIST-1h ought to be most ``stable'' planet amongst them, when viewed purely from the perspective of atmospheric ion loss. Along the same lines, it seems likely that TRAPPIST-1g will represent the best chance for a planet in the HZ of this planetary system to support a stable atmosphere over long periods.\footnote{3D climate simulations appear to suggest that TRAPPIST-1f and TRAPPIST-1g may not be amenable to surficial life as they enter a snowball state \citep{Wolf17}, but the effects of tidal heating, which are expected to be considerable \citep{Luger17}, were not included in the model.}

At this stage, we must reiterate the caveats discussed earlier. Firstly, most of the planetary and stellar parameters are partly or wholly unknown since the appropriate observations are not currently existent. The presence of (i) a more massive atmosphere, (ii) a different atmospheric composition, and (iii) the planet's magnetic field, are likely to alter the extent of atmospheric loss to some degree. Nonetheless, it can be surmised that the unmagnetized cases (with maximum total pressure) considered herein do yield robust upper bounds on the atmospheric ion escape rates. Apart from atmospheric loss, it is possible that outgassing processes could very well replenish the atmosphere \citep{KC03}. Hence, resolving the existence of an atmosphere over Gyr timescales necessitates an in-depth understanding of the interplay between source and loss mechanisms. Lastly, it is important to note that stellar properties evolve over time, implying that the escape rates are also likely to change accordingly. In the case of M-dwarfs such as TRAPPIST-1, the pre-main-sequence phase is particularly long and intense, and expected to have an adverse impact on atmospheric losses \citep{Luba15,SBJ16}. The ensuing effect of XUV on hydrodynamic and ion escape rates during this phase has not been investigated in the paper.

Bearing these limitations in mind, let us now turn to a discussion of the implications. We have argued that TRAPPIST-1h and TRAPPIST-1g represent the most promising candidates in terms of retaining atmospheres over Gyr timescales. Instead, if the atmosphere were to be depleted over $\mathcal{O}\left(10^8\right)$ years, this could prove to be problematic for the origin of life (abiogenesis) on the planet although it must be acknowledged that the actual timescale for abiogenesis on Earth and other planets remains unknown \citep{ST12}. Abiogenesis has been argued to be accompanied by an increase in biological (e.g. genomic) complexity over time \citep{AOC00,PuHe00} although this growth is not uniform, and may be contingent on environmental fluctuations. Hence, \emph{ceteris paribus}, a planet capable of sustaining a stable atmosphere over long time periods (along with retaining a stable climate) might have a greater chance of hosting complex surficial organisms. The outer planets of the TRAPPIST-1 system may therefore lead to more diverse biospheres eventually.

We have also shown that the ionospheric profiles for the TRAPPIST-1 planets in the HZ are not sensitive to the stellar wind conditions at altitudes $\lesssim 200$ km (see Fig. \ref{figIonProf} of the Appendix). This is an important result in light of the considerable variability and intensity of the stellar wind, since it suggests that the lower regions (such as the planetary surface) may remain mostly unaffected from under normal space weather conditions.\footnote{Stellar flares may have either a deleterious or beneficial effect on prebiotic chemistry that is dependent on a complex and interconnected set of factors \citep{Dart11,RWS17}.}

Let us turn our attention to Table \ref{tableesc} for the seven Earth-sized exoplanets of the TRAPPIST-1 system. It is seen that the ion escape rate reduces as one moves outwards. Hence, for similar multi-planetary systems around low-mass stars, it may be more prudent to focus on the outward planet(s) in the HZ for detecting atmospheres since their escape rates could be lower. Similarly, when confronted with two planets with similar values of $R_x$ and $a$ (see Eq. \ref{AnMassRat}), we propose that searches should focus on stars with lower mass-loss rates and stellar magnetic activity. Further details concerning the implications and generalizations of Eq. \ref{AnMassRat} will be reported in a forthcoming publication. Lastly, our results and implications are also broadly applicable to future planetary systems detected around M- and K-dwarfs endowed with similar features \citep{LL17}. 

To summarize, we have studied the atmospheric ion escape rates from the seven planets of the TRAPPIST-1 system by assuming a Venus-like composition. This was done by utilizing numerical models to compute the properties of the stellar wind and the escape rates, and the latter were shown to match the analytical predictions. We demonstrated that the outer planets of the TRAPPIST-1 system (most notable `h' and `g') are capable of retaining their atmospheres over Gyr timescales. However, as many factors remain unresolved at this stage, future missions such as the James Webb Space Telescope (JWST) will play a crucial role in constraining the atmospheres of the TRAPPIST-1 planets \citep{BI16}. In particular, a recent study concluded that spectral features for six of the seven TRAPPIST-1 planets could be detected with $< 20$ transits with 5 $\sigma$ accuracy \citep{MKR}. Such observations would help constrain theoretical predictions, pave the way towards looking for biosignatures, and empirically estimate the putative habitability of these planets.

\section*{Acknowledgements}
The authors thank A. Burrows, J.-F. Donati, S. Kane,  A. Knoll, A. Loeb and J. Winn for the helpful discussions and comments. The support provided by the NASA Living With a Star Jack Eddy Postdoctoral Fellowship Program, administered by the University Corporation for Atmospheric Research is acknowledged. M.J. is supported by NASA's SDO/AIA contract (NNG04EA00C) to LMSAL. Resources for this work were provided by the NASA High-End Computing (HEC) Program through the NASA Advanced Supercomputing (NAS) Division at Ames Research Center. The Space Weather Modeling Framework that contains the BATS-R-US code used in this study is publicly available from \url{http://csem.engin.umich.edu/tools/swmf}. For distribution of the model results used in this study, please contact the corresponding author.


\begin{thebibliography}{}

\bibitem[\protect\astroncite{{Adami} et~al.}{2000}]{AOC00}
{Adami}, C., {Ofria}, C., and {Collier}, T.~C. (2000).
\newblock {Evolution of biological complexity}.
\newblock {\em Proc. Natl. Acad. Sci. U.S.A.}, 97(9):4463--4468.

\bibitem[\protect\astroncite{{Airapetian} et~al.}{2017}]{Aira17}
{Airapetian}, V.~S., {Glocer}, A., {Khazanov}, G.~V., {Loyd}, R.~O.~P.,
  {France}, K., {Sojka}, J., {Danchi}, W.~C., and {Liemohn}, M.~W. (2017).
\newblock {How Hospitable Are Space Weather Affected Habitable Zones? The Role
  of Ion Escape}.
\newblock {\em Astrophys. J. Lett.}, 836:L3.

\bibitem[\protect\astroncite{{Airapetian} and {Usmanov}}{2016}]{AU16}
{Airapetian}, V.~S. and {Usmanov}, A.~V. (2016).
\newblock {Reconstructing the Solar Wind from Its Early History to Current
  Epoch}.
\newblock {\em Astrophys. J. Lett.}, 817(2):L24.

\bibitem[\protect\astroncite{{Angelo} et~al.}{2017}]{Ang17}
{Angelo}, I., {Rowe}, J.~F., {Howell}, S.~B., {Quintana}, E.~V., {Still}, M.,
  {Mann}, A.~W., {Burningham}, B., {Barclay}, T., {Ciardi}, D.~R., {Huber}, D.,
  and {Kane}, S.~R. (2017).
\newblock {Kepler-1649b: An Exo-Venus in the Solar Neighborhood}.
\newblock {\em Astron. J.}, 153(4):162.

\bibitem[\protect\astroncite{{Anglada-Escud{\'e}} et~al.}{2016}]{AE16}
{Anglada-Escud{\'e}}, G., {Amado}, P.~J., {Barnes}, J., {Berdi{\~n}as}, Z.~M.,
  {Butler}, R.~P., {Coleman}, G.~A.~L., {de La Cueva}, I., {Dreizler}, S.,
  {Endl}, M., {Giesers}, B., {Jeffers}, S.~V., {Jenkins}, J.~S., {Jones},
  H.~R.~A., {Kiraga}, M., {K{\"u}rster}, M., {L{\'o}pez-Gonz{\'a}lez}, M.~J.,
  {Marvin}, C.~J., {Morales}, N., {Morin}, J., {Nelson}, R.~P., {Ortiz}, J.~L.,
  {Ofir}, A., {Paardekooper}, S.-J., {Reiners}, A., {Rodr{\'{\i}}guez}, E.,
  {Rodr{\'{\i}}guez-L{\'o}pez}, C., {Sarmiento}, L.~F., {Strachan}, J.~P.,
  {Tsapras}, Y., {Tuomi}, M., and {Zechmeister}, M. (2016).
\newblock {A terrestrial planet candidate in a temperate orbit around Proxima
  Centauri}.
\newblock {\em Nature}, 536(7617):437--440.

\bibitem[\protect\astroncite{{Barstow} and {Irwin}}{2016}]{BI16}
{Barstow}, J.~K. and {Irwin}, P.~G.~J. (2016).
\newblock {Habitable worlds with JWST: transit spectroscopy of the TRAPPIST-1
  system?}
\newblock {\em Mon. Not. R. Astron. Soc.}, 461(1):L92--L96.

\bibitem[\protect\astroncite{{Bolmont} et~al.}{2017}]{Bel17}
{Bolmont}, E., {Selsis}, F., {Owen}, J.~E., {Ribas}, I., {Raymond}, S.~N.,
  {Leconte}, J., and {Gillon}, M. (2017).
\newblock {Water loss from terrestrial planets orbiting ultracool dwarfs:
  implications for the planets of TRAPPIST-1}.
\newblock {\em Mon. Not. R. Astron. Soc.}, 464(3):3728--3741.

\bibitem[\protect\astroncite{{Bourrier} et~al.}{2017}]{Bour17}
{Bourrier}, V., {Ehrenreich}, D., {Wheatley}, P.~J., {Bolmont}, E., {Gillon},
  M., {de Wit}, J., {Burgasser}, A.~J., {Jehin}, E., {Queloz}, D., and
  {Triaud}, A.~H.~M.~J. (2017).
\newblock {Reconnaissance of the TRAPPIST-1 exoplanet system in the
  Lyman-{$\alpha$} line}.
\newblock {\em Astron. Astrophys.}, 599:L3.

\bibitem[\protect\astroncite{{Brain} et~al.}{2016}]{brain16}
{Brain}, D.~A., {Bagenal}, F., {Ma}, Y.-J., {Nilsson}, H., and {Stenberg
  Wieser}, G. (2016).
\newblock Atmospheric escape from unmagnetized bodies.
\newblock {\em J. Geophys. Res. E}, 121(12):2364--2385.

\bibitem[\protect\astroncite{{Burkhart} and {Loeb}}{2017}]{BL17}
{Burkhart}, B. and {Loeb}, A. (2017).
\newblock {The Detectability of Radio Auroral Emission from Proxima b}.
\newblock {\em Astrophys. J. Lett.}, 849(1):L10.

\bibitem[\protect\astroncite{{Canto} and {Raga}}{1991}]{CR91}
{Canto}, J. and {Raga}, A.~C. (1991).
\newblock {Mixing layers in stellar outflows}.
\newblock {\em Astrophys. J.}, 372:646--658.

\bibitem[\protect\astroncite{{C{\'e}bron} and {Hollerbach}}{2014}]{CeHo14}
{C{\'e}bron}, D. and {Hollerbach}, R. (2014).
\newblock {Tidally Driven Dynamos in a Rotating Sphere}.
\newblock {\em Astrophys. J. Lett.}, 789(1):L25.

\bibitem[\protect\astroncite{{Chabrier}}{2003}]{Chab03}
{Chabrier}, G. (2003).
\newblock {Galactic Stellar and Substellar Initial Mass Function}.
\newblock {\em Publ. Astron. Soc. Pac.}, 115(809):763--795.

\bibitem[\protect\astroncite{{Christensen}}{2010}]{Christ10}
{Christensen}, U.~R. (2010).
\newblock {Dynamo Scaling Laws and Applications to the Planets}.
\newblock {\em Space Sci. Rev.}, 152(1-4):565--590.

\bibitem[\protect\astroncite{{Cockell} et~al.}{2016}]{Cock16}
{Cockell}, C.~S., {Bush}, T., {Bryce}, C., {Direito}, S., {Fox-Powell}, M.,
  {Harrison}, J.~P., {Lammer}, H., {Landenmark}, H., {Martin-Torres}, J.,
  {Nicholson}, N., {Noack}, L., {O'Malley-James}, J., {Payler}, S.~J.,
  {Rushby}, A., {Samuels}, T., {Schwendner}, P., {Wadsworth}, J., and
  {Zorzano}, M.~P. (2016).
\newblock {Habitability: A Review}.
\newblock {\em Astrobiology}, 16(1):89--117.

\bibitem[\protect\astroncite{{Cohen} et~al.}{2014}]{cohen14}
{Cohen}, O., {Drake}, J.~J., {Glocer}, A., {Garraffo}, C., {Poppenhaeger}, K.,
  {Bell}, J.~M., {Ridley}, A.~J., and {Gombosi}, T.~I. (2014).
\newblock {Magnetospheric Structure and Atmospheric Joule Heating of Habitable
  Planets Orbiting M-dwarf Stars}.
\newblock {\em Astrophys. J.}, 790:57.

\bibitem[\protect\astroncite{{Cuntz} et~al.}{2000}]{CSM00}
{Cuntz}, M., {Saar}, S.~H., and {Musielak}, Z.~E. (2000).
\newblock {On Stellar Activity Enhancement Due to Interactions with Extrasolar
  Giant Planets}.
\newblock {\em Astrophys. J. Lett.}, 533(2):L151--L154.

\bibitem[\protect\astroncite{{Dartnell}}{2011}]{Dart11}
{Dartnell}, L.~R. (2011).
\newblock {Ionizing Radiation and Life}.
\newblock {\em Astrobiology}, 11(6):551--582.

\bibitem[\protect\astroncite{{de Wit} et~al.}{2016}]{DeW17}
{de Wit}, J., {Wakeford}, H.~R., {Gillon}, M., {Lewis}, N.~K., {Valenti},
  J.~A., {Demory}, B.-O., {Burgasser}, A.~J., {Burdanov}, A., {Delrez}, L.,
  {Jehin}, E., {Lederer}, S.~M., {Queloz}, D., {Triaud}, A.~H.~M.~J., and {Van
  Grootel}, V. (2016).
\newblock {A combined transmission spectrum of the Earth-sized exoplanets
  TRAPPIST-1 b and c}.
\newblock {\em Nature}, 537(7618):69--72.

\bibitem[\protect\astroncite{{Dong} et~al.}{2017a}]{WW}
{Dong}, C., {Huang}, Z., {Lingam}, M., {T{\'o}th}, G., {Gombosi}, T., and
  {Bhattacharjee}, A. (2017a).
\newblock {The Dehydration of Water Worlds via Atmospheric Losses}.
\newblock {\em Astrophys. J. Lett.}, 847:L4.

\bibitem[\protect\astroncite{{Dong} et~al.}{2017b}]{DLMC}
{Dong}, C., {Lingam}, M., {Ma}, Y., and {Cohen}, O. (2017b).
\newblock {Is Proxima Centauri b Habitable? A Study of Atmospheric Loss}.
\newblock {\em Astrophys. J. Lett.}, 837(2):L26.

\bibitem[\protect\astroncite{{Dressing} and {Charbonneau}}{2015}]{DC15}
{Dressing}, C.~D. and {Charbonneau}, D. (2015).
\newblock {The Occurrence of Potentially Habitable Planets Orbiting M Dwarfs
  Estimated from the Full Kepler Dataset and an Empirical Measurement of the
  Detection Sensitivity}.
\newblock {\em Astrophys. J.}, 807:45.

\bibitem[\protect\astroncite{{Garcia-Sage} et~al.}{2017}]{Garcia-Sage17}
{Garcia-Sage}, K., {Glocer}, A., {Drake}, J.~J., {Gronoff}, G., and {Cohen}, O.
  (2017).
\newblock {On the Magnetic Protection of the Atmosphere of Proxima Centauri b}.
\newblock {\em Astrophys. J. Lett.}, 844:L13.

\bibitem[\protect\astroncite{{Garraffo} et~al.}{2016}]{GDC16}
{Garraffo}, C., {Drake}, J.~J., and {Cohen}, O. (2016).
\newblock {The Space Weather of Proxima Centauri b}.
\newblock {\em Astrophys. J. Lett.}, 833(1):L4.

\bibitem[\protect\astroncite{{Gillon} et~al.}{2016}]{Gill16}
{Gillon}, M., {Jehin}, E., {Lederer}, S.~M., {Delrez}, L., {de Wit}, J.,
  {Burdanov}, A., {Van Grootel}, V., {Burgasser}, A.~J., {Triaud}, A.~H.~M.~J.,
  {Opitom}, C., {Demory}, B.-O., {Sahu}, D.~K., {Bardalez Gagliuffi}, D.,
  {Magain}, P., and {Queloz}, D. (2016).
\newblock {Temperate Earth-sized planets transiting a nearby ultracool dwarf
  star}.
\newblock {\em Nature}, 533(7602):221--224.

\bibitem[\protect\astroncite{{Gillon} et~al.}{2017}]{Gill17}
{Gillon}, M., {Triaud}, A.~H.~M.~J., {Demory}, B.-O., {Jehin}, E., {Agol}, E.,
  {Deck}, K.~M., {Lederer}, S.~M., {de Wit}, J., {Burdanov}, A., {Ingalls},
  J.~G., {Bolmont}, E., {Leconte}, J., {Raymond}, S.~N., {Selsis}, F.,
  {Turbet}, M., {Barkaoui}, K., {Burgasser}, A., {Burleigh}, M.~R., {Carey},
  S.~J., {Chaushev}, A., {Copperwheat}, C.~M., {Delrez}, L., {Fernandes},
  C.~S., {Holdsworth}, D.~L., {Kotze}, E.~J., {Van Grootel}, V., {Almleaky},
  Y., {Benkhaldoun}, Z., {Magain}, P., and {Queloz}, D. (2017).
\newblock {Seven temperate terrestrial planets around the nearby ultracool
  dwarf star TRAPPIST-1}.
\newblock {\em Nature}, 542(7642):456--460.

\bibitem[\protect\astroncite{{Grie{\ss}meier}}{2015}]{Gri15}
{Grie{\ss}meier}, J.-M. (2015).
\newblock {Detection Methods and Relevance of Exoplanetary Magnetic Fields}.
\newblock In {Lammer}, H. and {Khodachenko}, M., editors, {\em Characterizing
  Stellar and Exoplanetary Environments}, volume 411 of {\em Astrophysics and
  Space Science Library}, page 213.

\bibitem[\protect\astroncite{{Grie{\ss}meier} et~al.}{2007}]{Grie07}
{Grie{\ss}meier}, J.-M., {Zarka}, P., and {Spreeuw}, H. (2007).
\newblock {Predicting low-frequency radio fluxes of known extrasolar planets}.
\newblock {\em Astron. Astrophys.}, 475:359--368.

\bibitem[\protect\astroncite{{Jakosky} et~al.}{2015}]{Jak15}
{Jakosky}, B.~M., {Grebowsky}, J.~M., {Luhmann}, J.~G., {Connerney}, J.,
  {Eparvier}, F., {Ergun}, R., {Halekas}, J., {Larson}, D., {Mahaffy}, P.,
  {McFadden}, J., {Mitchell}, D.~F., {Schneider}, N., {Zurek}, R., {Bougher},
  S., {Brain}, D., {Ma}, Y.~J., {Mazelle}, C., {Andersson}, L., {Andrews}, D.,
  {Baird}, D., {Baker}, D., {Bell}, J.~M., {Benna}, M., {Chaffin}, M.,
  {Chamberlin}, P., {Chaufray}, Y.-Y., {Clarke}, J., {Collinson}, G., {Combi},
  M., {Crary}, F., {Cravens}, T., {Crismani}, M., {Curry}, S., {Curtis}, D.,
  {Deighan}, J., {Delory}, G., {Dewey}, R., {DiBraccio}, G., {Dong}, C.,
  {Dong}, Y., {Dunn}, P., {Elrod}, M., {England}, S., {Eriksson}, A., {Espley},
  J., {Evans}, S., {Fang}, X., {Fillingim}, M., {Fortier}, K., {Fowler}, C.~M.,
  {Fox}, J., {Gr{\"o}ller}, H., {Guzewich}, S., {Hara}, T., {Harada}, Y.,
  {Holsclaw}, G., {Jain}, S.~K., {Jolitz}, R., {Leblanc}, F., {Lee}, C.~O.,
  {Lee}, Y., {Lefevre}, F., {Lillis}, R., {Livi}, R., {Lo}, D., {Mayyasi}, M.,
  {McClintock}, W., {McEnulty}, T., {Modolo}, R., {Montmessin}, F., {Morooka},
  M., {Nagy}, A., {Olsen}, K., {Peterson}, W., {Rahmati}, A., {Ruhunusiri}, S.,
  {Russell}, C.~T., {Sakai}, S., {Sauvaud}, J.-A., {Seki}, K., {Steckiewicz},
  M., {Stevens}, M., {Stewart}, A.~I.~F., {Stiepen}, A., {Stone}, S.,
  {Tenishev}, V., {Thiemann}, E., {Tolson}, R., {Toublanc}, D., {Vogt}, M.,
  {Weber}, T., {Withers}, P., {Woods}, T., and {Yelle}, R. (2015).
\newblock {MAVEN observations of the response of Mars to an interplanetary
  coronal mass ejection}.
\newblock {\em Science}, 350(6261):0210.

\bibitem[\protect\astroncite{{Jin} et~al.}{2012}]{jin12}
{Jin}, M., {Manchester}, W.~B., {van der Holst}, B., {Gruesbeck}, J.~R.,
  {Frazin}, R.~A., {Landi}, E., {Vasquez}, A.~M., {Lamy}, P.~L., {Llebaria},
  A., {Fedorov}, A., {Toth}, G., and {Gombosi}, T.~I. (2012).
\newblock {A Global Two-temperature Corona and Inner Heliosphere Model: A
  Comprehensive Validation Study}.
\newblock {\em Astrophys. J.}, 745:6.

\bibitem[\protect\astroncite{{Kasting} and {Catling}}{2003}]{KC03}
{Kasting}, J.~F. and {Catling}, D. (2003).
\newblock {Evolution of a Habitable Planet}.
\newblock {\em Annu. Rev. Astron. Astrophys.}, 41:429--463.

\bibitem[\protect\astroncite{{Khodachenko} et~al.}{2007}]{Khoda07}
{Khodachenko}, M.~L., {Ribas}, I., {Lammer}, H., {Grie{\ss}meier}, J.-M.,
  {Leitner}, M., {Selsis}, F., {Eiroa}, C., {Hanslmeier}, A., {Biernat}, H.~K.,
  {Farrugia}, C.~J., and {Rucker}, H.~O. (2007).
\newblock {Coronal Mass Ejection (CME) Activity of Low Mass M Stars as An
  Important Factor for The Habitability of Terrestrial Exoplanets. I. CME
  Impact on Expected Magnetospheres of Earth-Like Exoplanets in Close-In
  Habitable Zones}.
\newblock {\em Astrobiology}, 7(1):167--184.

\bibitem[\protect\astroncite{{Kopparapu} et~al.}{2013}]{Kop13}
{Kopparapu}, R.~K., {Ramirez}, R., {Kasting}, J.~F., {Eymet}, V., {Robinson},
  T.~D., {Mahadevan}, S., {Terrien}, R.~C., {Domagal-Goldman}, S., {Meadows},
  V., and {Deshpande}, R. (2013).
\newblock {Habitable Zones around Main-sequence Stars: New Estimates}.
\newblock {\em Astrophys. J.}, 765(2):131.

\bibitem[\protect\astroncite{{Lammer}}{2013}]{Lammer13}
{Lammer}, H. (2013).
\newblock {\em {Origin and Evolution of Planetary Atmospheres: Implications for
  Habitability}}.
\newblock Springer Briefs in Astronomy. Springer.

\bibitem[\protect\astroncite{{Lammer} et~al.}{2009}]{Lam09}
{Lammer}, H., {Bredeh{\"o}ft}, J.~H., {Coustenis}, A., {Khodachenko}, M.~L.,
  {Kaltenegger}, L., {Grasset}, O., {Prieur}, D., {Raulin}, F., {Ehrenfreund},
  P., {Yamauchi}, M., {Wahlund}, J.-E., {Grie{\ss}meier}, J.-M., {Stangl}, G.,
  {Cockell}, C.~S., {Kulikov}, Y.~N., {Grenfell}, J.~L., and {Rauer}, H.
  (2009).
\newblock {What makes a planet habitable?}
\newblock {\em Astron. Astrophys. Rev.}, 17(2):181--249.

\bibitem[\protect\astroncite{{Lingam} and {Loeb}}{2017a}]{LL17}
{Lingam}, M. and {Loeb}, A. (2017a).
\newblock {Enhanced interplanetary panspermia in the TRAPPIST-1 system}.
\newblock {\em Proc. Natl. Acad. Sci. U.S.A.}, 114(26):6689--6693.

\bibitem[\protect\astroncite{{Lingam} and {Loeb}}{2017b}]{LiLo17}
{Lingam}, M. and {Loeb}, A. (2017b).
\newblock {Physical constraints on the likelihood of life on exoplanets}.
\newblock {\em Int. J. Astrobiol. (arXiv:1707.02996)}.

\bibitem[\protect\astroncite{{Luger} and {Barnes}}{2015}]{Luba15}
{Luger}, R. and {Barnes}, R. (2015).
\newblock {Extreme Water Loss and Abiotic O2Buildup on Planets Throughout the
  Habitable Zones of M Dwarfs}.
\newblock {\em Astrobiology}, 15(2):119--143.

\bibitem[\protect\astroncite{{Luger} et~al.}{2017}]{Luger17}
{Luger}, R., {Sestovic}, M., {Kruse}, E., {Grimm}, S.~L., {Demory}, B.-O.,
  {Agol}, E., {Bolmont}, E., {Fabrycky}, D., {Fernandes}, C.~S., {Van Grootel},
  V., {Burgasser}, A., {Gillon}, M., {Ingalls}, J.~G., {Jehin}, E., {Raymond},
  S.~N., {Selsis}, F., {Triaud}, A.~H.~M.~J., {Barclay}, T., {Barentsen}, G.,
  {Howell}, S.~B., {Delrez}, L., {de Wit}, J., {Foreman-Mackey}, D.,
  {Holdsworth}, D.~L., {Leconte}, J., {Lederer}, S., {Turbet}, M., {Almleaky},
  Y., {Benkhaldoun}, Z., {Magain}, P., {Morris}, B.~M., {Heng}, K., and
  {Queloz}, D. (2017).
\newblock {A seven-planet resonant chain in TRAPPIST-1}.
\newblock {\em Nat. Astron.}, 1:0129.

\bibitem[\protect\astroncite{{Ma} et~al.}{2004}]{Ma04}
{Ma}, Y., {Nagy}, A.~F., {Sokolov}, I.~V., and {Hansen}, K.~C. (2004).
\newblock {Three-dimensional, multispecies, high spatial resolution MHD studies
  of the solar wind interaction with Mars}.
\newblock {\em J. Geophys. Res. A}, 109:A07211.

\bibitem[\protect\astroncite{{Ma} et~al.}{2013}]{Ma13}
{Ma}, Y.~J., {Nagy}, A.~F., {Russell}, C.~T., {Strangeway}, R.~J., {Wei},
  H.~Y., and {Toth}, G. (2013).
\newblock {A global multispecies single-fluid MHD study of the plasma
  interaction around Venus}.
\newblock {\em J. Geophys. Res. A}, 118:321--330.

\bibitem[\protect\astroncite{{Morin} et~al.}{2010}]{morin10}
{Morin}, J., {Donati}, J.-F., {Petit}, P., {Delfosse}, X., {Forveille}, T., and
  {Jardine}, M.~M. (2010).
\newblock {Large-scale magnetic topologies of late M dwarfs}.
\newblock {\em Mon. Not. R. Astron. Soc.}, 407(4):2269--2286.

\bibitem[\protect\astroncite{{Morley} et~al.}{2017}]{MKR}
{Morley}, C.~V., {Kreidberg}, L., {Rustamkulov}, Z., {Robinson}, T., and
  {Fortney}, J.~J. (2017).
\newblock {Observing the Atmospheres of Known Temperate Earth-sized Planets
  with JWST}.
\newblock {\em Astrophys. J.}, 850:121.

\bibitem[\protect\astroncite{{Oran} et~al.}{2013}]{oran13}
{Oran}, R., {van der Holst}, B., {Landi}, E., {Jin}, M., {Sokolov}, I.~V., and
  {Gombosi}, T.~I. (2013).
\newblock {A Global Wave-driven Magnetohydrodynamic Solar Model with a Unified
  Treatment of Open and Closed Magnetic Field Topologies}.
\newblock {\em Astrophys. J.}, 778:176.

\bibitem[\protect\astroncite{{Owen} and {Alvarez}}{2016}]{OA16}
{Owen}, J.~E. and {Alvarez}, M.~A. (2016).
\newblock {UV Driven Evaporation of Close-in Planets: Energy-limited,
  Recombination-limited, and Photon-limited Flows}.
\newblock {\em Astrophys. J.}, 816(1):34.

\bibitem[\protect\astroncite{{Pont}}{2009}]{Pont09}
{Pont}, F. (2009).
\newblock {Empirical evidence for tidal evolution in transiting planetary
  systems}.
\newblock {\em Mon. Not. R. Astron. Soc.}, 396(3):1789--1796.

\bibitem[\protect\astroncite{{Purvis} and {Hector}}{2000}]{PuHe00}
{Purvis}, A. and {Hector}, A. (2000).
\newblock Getting the measure of biodiversity.
\newblock {\em Nature}, 405(6783):212--219.

\bibitem[\protect\astroncite{{Ranjan} et~al.}{2017}]{RWS17}
{Ranjan}, S., {Wordsworth}, R., and {Sasselov}, D.~D. (2017).
\newblock {The Surface UV Environment on Planets Orbiting M-Dwarfs:
  Implications for Prebiotic Chemistry and the Need for Experimental
  Follow-up}.
\newblock {\em Astrophys. J.}, 843(2):110.

\bibitem[\protect\astroncite{{Shields} et~al.}{2016}]{SBJ16}
{Shields}, A.~L., {Ballard}, S., and {Johnson}, J.~A. (2016).
\newblock {The habitability of planets orbiting M-dwarf stars}.
\newblock {\em Phys. Rep.}, 663(1):1--38.

\bibitem[\protect\astroncite{{Smith} and {Smith}}{1972}]{smith1972}
{Smith}, F.~L. and {Smith}, C. (1972).
\newblock {Numerical evaluation of Chapman's grazing incidence integral ch (X,
  {$\chi$})}.
\newblock {\em \jgr}, 77:3592--3597.

\bibitem[\protect\astroncite{{Sokolov} et~al.}{2013}]{SVO13}
{Sokolov}, I.~V., {van der Holst}, B., {Oran}, R., {Downs}, C., {Roussev},
  I.~I., {Jin}, M., {Manchester}, IV, W.~B., {Evans}, R.~M., and {Gombosi},
  T.~I. (2013).
\newblock {Magnetohydrodynamic Waves and Coronal Heating: Unifying Empirical
  and MHD Turbulence Models}.
\newblock {\em Astrophys. J.}, 764(1):23.

\bibitem[\protect\astroncite{{Spiegel} and {Turner}}{2012}]{ST12}
{Spiegel}, D.~S. and {Turner}, E.~L. (2012).
\newblock {Bayesian analysis of the astrobiological implications of life's
  early emergence on Earth}.
\newblock {\em Proc. Natl. Acad. Sci.}, 109(2):395--400.

\bibitem[\protect\astroncite{{T{\'o}th} et~al.}{2011}]{TVH11}
{T{\'o}th}, G., {van der Holst}, B., and {Huang}, Z. (2011).
\newblock {Obtaining Potential Field Solutions with Spherical Harmonics and
  Finite Differences}.
\newblock {\em Astrophys. J.}, 732(2):102.

\bibitem[\protect\astroncite{{T{\'o}th} et~al.}{2012}]{Toth12}
{T{\'o}th}, G., {van der Holst}, B., {Sokolov}, I.~V., {De Zeeuw}, D.~L.,
  {Gombosi}, T.~I., {Fang}, F., {Manchester}, W.~B., {Meng}, X., {Najib}, D.,
  {Powell}, K.~G., {Stout}, Q.~F., {Glocer}, A., {Ma}, Y.-J., and {Opher}, M.
  (2012).
\newblock {Adaptive numerical algorithms in space weather modeling}.
\newblock {\em J. Comp. Phys.}, 231(3):870--903.

\bibitem[\protect\astroncite{{Tripathi} et~al.}{2015}]{Tripathi2015}
{Tripathi}, A., {Kratter}, K.~M., {Murray-Clay}, R.~A., and {Krumholz}, M.~R.
  (2015).
\newblock {Simulated Photoevaporative Mass Loss from Hot Jupiters in 3D}.
\newblock {\em \apj}, 808:173.

\bibitem[\protect\astroncite{{van der Holst} et~al.}{2014}]{bart14}
{van der Holst}, B., {Sokolov}, I.~V., {Meng}, X., {Jin}, M., {Manchester}, IV,
  W.~B., {T{\'o}th}, G., and {Gombosi}, T.~I. (2014).
\newblock {Alfv{\'e}n Wave Solar Model (AWSoM): Coronal Heating}.
\newblock {\em Astrophys. J.}, 782:81.

\bibitem[\protect\astroncite{{Vidotto} and {Donati}}{2017}]{VD17}
{Vidotto}, A.~A. and {Donati}, J.-F. (2017).
\newblock {Predicting radio emission from the newborn hot Jupiter V830 Tauri b
  and its host star}.
\newblock {\em Astron. Astrophys.}, 602:A39.

\bibitem[\protect\astroncite{{Vidotto} et~al.}{2013}]{vidotto13}
{Vidotto}, A.~A., {Jardine}, M., {Morin}, J., {Donati}, J.-F., {Lang}, P., and
  {Russell}, A.~J.~B. (2013).
\newblock {Effects of M dwarf magnetic fields on potentially habitable
  planets}.
\newblock {\em Astron. Astrophys.}, 557:A67.

\bibitem[\protect\astroncite{{Wheatley} et~al.}{2017}]{Wheat17}
{Wheatley}, P.~J., {Louden}, T., {Bourrier}, V., {Ehrenreich}, D., and
  {Gillon}, M. (2017).
\newblock {Strong XUV irradiation of the Earth-sized exoplanets orbiting the
  ultracool dwarf TRAPPIST-1}.
\newblock {\em Mon. Not. R. Astron. Soc. Lett.}, 465(1):L74--L78.

\bibitem[\protect\astroncite{{Winn} and {Fabrycky}}{2015}]{WF15}
{Winn}, J.~N. and {Fabrycky}, D.~C. (2015).
\newblock {The Occurrence and Architecture of Exoplanetary Systems}.
\newblock {\em Annu. Rev. Astron. Astrophys.}, 53:409--447.

\bibitem[\protect\astroncite{{Wolf}}{2017}]{Wolf17}
{Wolf}, E.~T. (2017).
\newblock {Assessing the Habitability of the TRAPPIST-1 System Using a 3D
  Climate Model}.
\newblock {\em Astrophys. J. Lett.}, 839(1):L1.

\bibitem[\protect\astroncite{{Wood} et~al.}{2001}]{WLMZ}
{Wood}, B.~E., {Linsky}, J.~L., {M{\"u}ller}, H.-R., and {Zank}, G.~P. (2001).
\newblock {Observational Estimates for the Mass-Loss Rates of {$\alpha$}
  Centauri and Proxima Centauri Using Hubble Space Telescope Ly{$\alpha$}
  Spectra}.
\newblock {\em Astrophys. J. Lett.}, 547(1):L49--L52.

\bibitem[\protect\astroncite{{Yantis} et~al.}{1977}]{YSE77}
{Yantis}, W.~F., {Sullivan}, III, W.~T., and {Erickson}, W.~C. (1977).
\newblock {A Search for Extra-Solar Jovian Planets by Radio Techniques}.
\newblock In {\em Bulletin of the American Astronomical Society}, volume~9 of
  {\em \baas}, page 453.

\bibitem[\protect\astroncite{{Zahnle} and {Catling}}{2017}]{ZC17}
{Zahnle}, K.~J. and {Catling}, D.~C. (2017).
\newblock {The Cosmic Shoreline: The Evidence that Escape Determines which
  Planets Have Atmospheres, and what this May Mean for Proxima Centauri B}.
\newblock {\em Astrophys. J.}, 843(2):122.

\bibitem[\protect\astroncite{{Zarka}}{1998}]{Zark98}
{Zarka}, P. (1998).
\newblock {Auroral radio emissions at the outer planets: Observations and
  theories}.
\newblock {\em J. Geophys. Res.}, 103(E9):20159--20194.

\bibitem[\protect\astroncite{{Zarka}}{2007}]{Zark07}
{Zarka}, P. (2007).
\newblock {Plasma interactions of exoplanets with their parent star and
  associated radio emissions}.
\newblock {\em Planet. Space Sci.}, 55(5):598--617.

\bibitem[\protect\astroncite{{Zendejas} et~al.}{2010}]{ZSR10}
{Zendejas}, J., {Segura}, A., and {Raga}, A.~C. (2010).
\newblock {Atmospheric mass loss by stellar wind from planets around main
  sequence M stars}.
\newblock {\em Icarus}, 210(2):539--544.

\bibitem[\protect\astroncite{{Zsom}}{2015}]{Zs15}
{Zsom}, A. (2015).
\newblock {A Population-based Habitable Zone Perspective}.
\newblock {\em Astrophys. J.}, 813(1):9.

\end{thebibliography}

\appendix

\section{Modelling the stellar wind of TRAPPIST-1} \label{AppA}
In this Section, we describe the methodology involved in modelling the stellar wind of TRAPPIST-1 in more detail.

\subsection{Description of the Alfv\'{e}n Wave Solar Model}
In this study, the TRAPPIST-1 stellar wind is simulated by means of the Alfv\'{e}n Wave Solar Model (AWSoM; \citealp{bart14}), a data-driven global MHD model that had been initially developed for simulating the solar atmosphere and solar wind. The inner boundary condition of the magnetic field can be specified by different magnetic maps from available observations. The initial conditions for the stellar wind plasma are determined through the Parker solution, while the initial magnetic field is based on the Potential Field Source Surface (PFSS) model with the Finite Difference Iterative Potential Solver (FDIPS) described in \citet{TVH11}. 

Alfv\'{e}n waves are driven at the inner boundary with a Poynting flux that scales with the surface magnetic field. The stellar wind is heated by Alfv\'{e}n wave dissipation and accelerated by thermal and Alfv\'{e}n wave pressure. Electron heat conduction (that includes both collisional and collisionless contributions) and radiative cooling are also included in the model. In the AWSoM, the electron and proton temperatures are treated separately, while the electrons and protons are assumed to have the same bulk velocity. However, heat conduction is applied only to the electrons, owing to their much higher thermal velocity. The system of governing equations is solved numerically using the Block Adaptive Tree Solar Wind Roe-type Upwind Scheme (BATS-R-US) code within the Space Weather Modeling Framework \citep{Toth12}.

By using a physically consistent treatment of wave reflection, dissipation, and heat partitioning between the electrons and protons, the AWSoM has successfully reproduced solar coronal conditions to a high degree of precision.

\subsection{Application of the AWSoM to TRAPPIST-1}
To adapt the AWSoM for the TRAPPIST-1, we modify the rotational mass, radius, and period of the star in accordance with the latest observational data \citep{Luger17}. Hence, we specify $M_{\star}=0.08 M_\odot$, $R_{\star}=0.11 R_\odot$, and $P_{\star}=3.3$ days. 

Due to the lack of direct surface magnetic field observations of TRAPPIST-1, we utilize a solar magnetogram under the solar minimum condition (GONG magnetogram of Carrington Rotation 2077; \citealp{jin12}) and scale the mean and radial magnetic field strength based on the magnetic field observations of similar late M-dwarfs \citep{morin10}. Based on the fact that the X-ray luminosity of TRAPPIST-1 is similar to that of the quiet Sun \citep{Bour17}, we modify the Poynting flux parameter of the model such that the same amount of Poynting flux is generated as in the solar case. The simulation domain is extended to 250 $R_{\star}$ to ensure that the orbits of all seven planets in the TRAPPIST-1 system are duly encompassed. 

\section{Deducing the magnetic fields of the TRAPPIST-1 planets} \label{AppB}
Since the past 40 years, it has been well known that radio emission from exoplanets can be used to detect them, and thereby determine their magnetic fields \citep{YSE77}. The basic idea behind detecting the planetary magnetic field ($B_p$) is that the cyclotron maser instability drives the emission of radio waves at a frequency approximately equal to the gyrofrequency \citep{Zark98}. The maximum emission frequency $\omega$ is given by
\begin{equation}
    \omega \approx \omega_c = \frac{e B}{m_e c},
\end{equation}
where $m_e$ and $e$ are the electron mass and charge respectively. Hence, we propose that the radio auroral emission from the TRAPPIST-1 planets may be detectable by ground-based observatories when the frequency is above $\sim 10$ MHz; the Earth's ionosphere reflects radio waves below this value thereby requiring space- or lunar-based telescopes \citep{Gri15}. As an illustrative example, let us suppose that $B_p \approx 0.1$ G for the TRAPPIST-1 planets, which leads us to $f = \omega/2\pi \approx 0.3$ MHz. This fiducial value is motivated by planetary dynamo scaling laws \citep{Christ10}, as well as the lower limit of the hypothesized magnetic field for Proxima b \citep{BL17}.

Next, it is necessary to obtain a heuristic estimate of the radio flux density $\Phi$ at Earth to know whether the emission would be detectable. It can be estimated via 
\begin{equation} \label{RFD}
    \Phi = \frac{P_\mathrm{radio}}{4\pi \Delta{f} D^2},
\end{equation}
where bandwidth $\Delta{f} \approx f/2$, $D$ is the distance to Earth and $P_\mathrm{radio}$ represents the planetary radio power \citep{Zark07}. Thus, a knowledge of $P_\mathrm{radio}$ and $\omega$ would suffice to determine $\Phi$. Although there exist several ways of computing $P_\mathrm{radio}$ \citep{Gri15}, it has been noted that the planetary radio power is dominated by the dissipation of the magnetic power carried by the stellar wind \citep{VD17}:
\begin{equation} \label{RadP}
    P_\mathrm{radio} \sim 2 \times 10^{-3}\,\left(\frac{\pi B_{sw}^2 r_m^2 V_{sw}}{4\pi}\right),
\end{equation}
where $B_{sw}$ is the interplanetary magnetic field (IMF) and $V_{sw}$ is the stellar wind velocity, and both of them have been listed in Table \ref{tableSW}. Here, $r_m$ denotes the planetary magnetospheric radius and is given by
\begin{equation} \label{Mag}
    r_m = R_p \left(\frac{B_p^2}{8\pi P_{sw}}\right)^{1/6},
\end{equation}
where $P_{sw} \approx \rho_{sw} v_{sw}^2$ is the dynamic pressure and $\rho_{sw}$ is the stellar wind density provided in Table \ref{tableSW}. Hence, it is possible to combine Eqs. (\ref{RFD}), (\ref{RadP}) and (\ref{Mag}) to arrive at an estimate of $\Phi$. We find that the radio flux density, for the above choice of values, is $\mathcal{O}\left(10^{-4}\right)$ Jy. However, we wish to note that $\Phi$ can be boosted by a factor of $\sim 10^2-10^3$ during a strong Coronal Mass Ejection (CME) event, which would imply that it can attain a value of $\sim 10-100$ mJy. This follows from the fact that the stellar wind parameters are enhanced during a CME event, and the corresponding radio power is also increased accordingly \citep{Grie07}. We note that a similar approach has also been employed in the context of Proxima b, where it was shown that the radio flux density attained a peak value of $\sim 1-10$ Jy during a large Carrington-type CME event.

Thus, to summarize, future space-based (or lunar) low-frequency observations may be able to constrain the planetary magnetic fields of the TRAPPIST-1 planets by measuring the radio flux density and extrapolating backwards to determine the value of $B_p$.

\section{Atmospheric ion escape rates for the TRAPPIST-1 planets}
In this Section, we briefly describe the workings of the code and provide additional results pertaining to the atmospheric ion escape rates.

\subsection{Physical model and computational methodology}
The 3-D Block Adaptive Tree Solar-wind Roe Upwind Scheme (BATS-R-US) multi-species MHD (MS-MHD) model was initially developed in the context of our Solar system i.e. for studying Mars \citep{Ma04} and Venus \citep{Ma13}. We rely upon the code developed for Venus, and the neutral atmospheric profiles are based on the solar maximum conditions. The MS-MHD comprises of a separate continuity equation for each ion species, in conjunction with one momentum equation and one energy equation for the four ion fluids H$^+$, O$^+$, O$_2^+$, CO$_2^+$ \citep{Ma04,Ma13}. Unlike most global (Earth) magnetosphere models that commence from $2$-$3$ Earth radii, the Mars/Venus MS-MHD model contains a \emph{self-consistent} ionosphere, and thus the lower boundary extends down to an altitude of 100 km above the planetary surface. The MS-MHD model, which serves as the basis of our paper, accounts for a diverse array of chemical processes, such as charge exchange, photoionization and electron recombination.

The various chemical reactions and their corresponding rate coefficients have been delineated in Table \ref{tableCM}, and the reader may also consult \citet{DLMC} for further details. The densities of O$^+$, O$_2^+$, CO$_2^+$ at the lower boundary satisfy the photochemical equilibrium condition as described in \citet{DLMC}. The model also assumes that the plasma temperature (sum of ion and electron temperatures) is approximately double that of the neutral temperature at the lower boundary because of the high collision frequency.

The grid is also taken to be non-uniform and spherical in nature to accurately capture the multi-scale physics operating in different regions. Hence, the radial resolution ranges from around half the scale height at the lower boundary to several thousands of kilometers at the outer boundary. The horizontal resolution is chosen to be $3.0^{\circ}$ (in both longitude and latitude), while the simulation domain ranges from $-$45 R$_x \leq$ X $\leq$ 15 R$_x$ and $-$30 R$_x \leq$ Y, Z $\leq$ 30 R$_x$, where R$_x$ denotes the radius of planet $X$. The code is run in the Planet-Star-Orbital (PSO) coordinate system, where the $x$-axis is directed from the planet towards TRAPPIST-1, the $z$-axis is perpendicular to the planet's orbital plane, and the $y$-axis completes the right-hand system.

\begin{table*}[!h]
\centering
\caption{Chemical reactions and associated rates adapted from \citet{DLMC}. The photoionization frequencies are rescaled to account for the EUV fluxes received at each of the TRAPPIST-1 planets based on the estimates provided in \citet{Bel17} and \citet{Bour17}.}\label{tableCM}
\begin{tabular}{lll}
\hline
\hline
\multicolumn{2}{c} {Chemical Reaction} & Rate Coefficient \\
\hline
\multicolumn{3}{c} {Primary Photolysis in s$^{-1}$} \\
\hline
& CO$_{2}$ + $h\nu$ $\rightarrow$  CO$_2^+$ + $e^{-}$ &  see table caption   \\ 
& O + $h\nu$ $\rightarrow$  O$^+$ + $e^{-}$     &  see table caption  \\
\hline
\multicolumn{3}{c} {Ion-neutral Chemistry in cm$^{3}$ s$^{-1}$} \\
\hline
&  CO$_{2}^+$ + O $\rightarrow$  O$_2^+$ + CO  & $1.64 \times 10^{-10} $   \\
&  CO$_{2}^+$ + O $\rightarrow$  O$^+$ + CO$_2$  & $9.60 \times 10^{-11} $   \\
&  O$^{+}$ + CO$_2$ $\rightarrow$  O$_2^+$ + CO  & $1.1 \times 10^{-9}$ (800/T$_i$)$^{0.39}$   \\
&  H$^{+}$ + O $\rightarrow$  O$^+$ + H  & $5.08 \times 10^{-10} $   \\
\hline
\multicolumn{3}{c} {Electron Recombination Chemistry in cm$^{3}$s$^{-1}$} \\
\hline
&  O$_2^{+}$ + $e^{-}$ $\rightarrow$  O + O  & $7.38 \times 10^{-8}$ (1200/T$_e$)$^{0.56}$   \\
&  CO$_2^{+}$ + $e^{-}$ $\rightarrow$  CO + O  & $3.10 \times 10^{-7}$ (300/T$_e$)$^{0.5}$  \\
\hline
\hline
\end{tabular}
\end{table*}

The multi-species MHD equations are summarized as follows:
\begin{eqnarray} \label{ContEq}
&& \frac{\partial \rho_{s}}{\partial t}+\nabla\cdot\left(\rho_{s}\mathbf{u}\right)=\mathcal{S}_{s}-\mathcal{L}_{s} \\ \label{MomEq}
&& \frac{\partial (\rho\mathbf{u})}{\partial t}+\nabla\cdot\left(\rho\mathbf{uu}+p\mathbf{I}+\frac{B^2}{2\mu_0}\mathbf{I}-\frac{1}{\mu_0}\mathbf{BB}\right)  =\rho\mathbf{G}
- \sum_{s=ions}\rho_{s}\sum_{n=neutrals}\nu_{sn}\mathbf{u}-\sum_{s=ions}\mathcal{L}_{s}\mathbf{u} \\ \label{EnEq}
&&\frac{\partial}{\partial t}\left(\frac{\rho u^2}{2} + \frac{p}{\ga-1}+\frac{B^2}{2\mu_0}\right)
+\nabla\cdot \left[\left(\frac{\rho u^2}{2} + \frac{\ga p}{\ga-1}+\frac{B^2}{\mu_0}\right)\mathbf{u}-\frac{(\mathbf{B}\cdot
\mathbf{u})\mathbf{B}}{\mu_0}+\mathbf{B}\times\frac{\nabla\times\mathbf{B}}{\mu_0^2\sigma_0}\right]\\ \nonumber
&& = \sum_{s=ions}\sum_{n=neutrals}\frac{\rho_{s}\nu_{sn}}{m_{s}+m_{n}}\left[3k\left(T_{n}-T_{s}\right)-m_{s}u^{2}\right] +\rho \mathbf{u} \cdot \mathbf{G}  \\ \nonumber
&& + \frac{k}{\gamma-1}\sum_{s=ions}\frac{\left(\mathcal{S}_{s}T_{n}-\mathcal{L}_{s}T_{s}\right)}{m_{s}} 
-\frac{1}{2}\sum_{s=ions}\mathcal{L}_{s}u^2+\frac{k}{\gamma-1}\frac{\mathcal{S}_{e}T_{n0}-\mathcal{L}_{e}T_{e}}{m_{e}} \\ \label{IndEq}
&& \frac{\mathbf{\partial B}}{\partial t}+\nabla\cdot\left(\mathbf{u}\mathbf{B}-\mathbf{B}\mathbf{u}\right)
=-\nabla\times\left(\frac{1}{\mu_0\sigma_0} \nabla\times\mathbf{B}\right) 
\end{eqnarray}
where Eqs. (\ref{ContEq}), (\ref{MomEq}) and (\ref{EnEq}) represent the conservation of mass (of each species), momentum and energy respectively, whilst Eq. (\ref{IndEq}) is the magnetic induction equation. The total mass density $\rho=\sum_{s=ions}\rho_s$ and $\nu_{sn}$ is the elastic ion-neutral collision frequency \citep{Ma04,Ma13}. The ratio of the specific heats $\ga$ is taken to be 5/3. The subscripts $s$, $n$ and $e$ indicate ion species $s$, neutral species $n$ and electron $e$. The other symbols have their usual definitions \citep{Ma04,Ma13}. In order to account for photoionization, we calculate the optical depth of the neutral atmosphere by applying the Chapman functions based on the numerical evaluation given by \citet{smith1972}. The photoelectron gains an excess energy through the photoionization process, as indicated by the presence of $T_{n0}$ in Eq. (\ref{EnEq}). Therefore, we include the stellar heating via photoionization as those hydrodynamic models \citep[e.g.,][]{Tripathi2015}. In the above set of equations, note that the source ($\mathcal{S}$) and loss ($\mathcal{L}$) terms of species $s$ associated with photoionization ($\nu_{ph,s^{\prime}}$), charge exchange ($k_{is^{\prime}}$) and recombination ($\alpha_{R,s}$) are shown as follows:
\begin{eqnarray}
\mathcal{S}_{s}&=&m_{s}n_{s^{\prime}}\left(\nu_{ph,s^{\prime}}+\sum_{i=ions} k_{is^{\prime}}n_{i}\right) \\
\mathcal{L}_{s}&=&m_{s}n_{s}\left(\alpha_{R,s}n_{e}+\sum_{n^{\prime}=neutrals} k_{sn^{\prime}}n_{n\prime}\right) \\
\mathcal{S}_{e}&=&m_{e}\sum_{s'=neutrals}\nu_{ph,s^{\prime}} n_{s'} \\
\mathcal{L}_{e}&=&m_{e}n_{e}\sum_{s=ions}\alpha_{R,s}n_{s}
\end{eqnarray}

In Eq. (\ref{IndEq}), $\sigma_0$ is the electrical conductivity. In the model, the electrical conductivity in the planetary ionosphere is calculated using:
\begin{eqnarray} \label{s1}
\sigma_0 = \frac{n_e e^2}{m_e(\nu_{ei}+\nu_{en})},
\end{eqnarray}
where m$_e$ is the electron mass, $\nu_{ei}$ and $\nu_{en}$ are the electron-ion and electron-neutral collision frequencies, respectively. The collision frequency $\nu_{ei}$ and $\nu_{en}$ are given by \citet{Ma13}
\begin{eqnarray} \label{s2}
\nu_{ei}=54.5\frac{n_i}{T_e^{3/2}},
\end{eqnarray}
\begin{eqnarray} \label{s3}
\nu_{en}&=&3.68\times10^{-8}\left(1+4.1\times10^{-11}|4500-T_e|^{2.93} \right)[CO_2] \\ \nonumber
&&+8.9\times10^{-11}(1+5.7\times10^{-4}T_e)T_e^{1/2}[O]
\end{eqnarray}
The above set of equations is solved using an upwind finite-volume scheme based on an approximate Riemann solver, to ensure the appropriate conservation of plasma variables. To determine the steady-state solution, the simulation starts with a 2-stage local-time stepping scheme that enables different grid cells to select different advance time step thus accelerating convergence to the steady-state to save computational resources. Because of the stiffness of the source terms, a point implicit scheme is used for handling them. 

\subsection{Auxiliary results} \label{SSecAux}
In the main paper, the atmospheric ion escape rates were provided for two cases: (i) maximum dynamic and total pressure, and (ii) minimum total pressure, but maximum magnetic pressure. The corresponding stellar wind parameters for these two cases have been delineated in Table \ref{tableSW}.

\begin{table}
\caption{Stellar wind parameters for the TRAPPIST-1 planets}\label{tableSW}
\centering
\begin{tabular}{lllll}
\hline
& N\textsubscript {sw}\par
(cm\textsuperscript {-3}) & T\textsubscript {sw}\par
(K) & V\textsubscript {sw}\par
(km/s) & IMF\par
(nT)  \\
\hline
\multicolumn{5}{c} {Maximum total pressure} \\
\hline
Trappist-1b & 6.59$\times$10$^{4}$  & 2.01$\times$10$^{6}$   &  $\left(-470,80,-1\right)$   &  $\left(381,81,-147\right)$ \\ 
Trappist-1c & 2.99$\times$10$^{4}$  & 1.68$\times$10$^{6}$   &  $\left(-527,68,0\right)$   &  $\left(210,42,-111\right)$ \\ 
Trappist-1d & 1.20$\times$10$^{4}$  & 1.59$\times$10$^{6}$   &  $\left(-566,56,6\right)$   &  $\left(-129,15,-50\right)$ \\
Trappist-1e & 5.79$\times$10$^{3}$  & 1.26$\times$10$^{6}$   &  $\left(-604,50,3\right)$   &  $\left(-149,13,-42\right)$ \\ 
Trappist-1f & 2.99$\times$10$^{3}$  & 1.02$\times$10$^{6}$   &  $\left(-624,44,3\right)$   &  $\left(-98,7,-34\right)$ \\ 
Trappist-1g & 1.95$\times$10$^{3}$  & 8.92$\times$10$^{5}$   &  $\left(-637,40,2\right)$   &  $\left(-69,6,-28\right)$ \\ 
Trappist-1h & 9.52$\times$10$^{2}$  & 7.17$\times$10$^{5}$   &  $\left(-657,37,2\right)$   &  $\left(-44,2,-25\right)$ \\ 
\hline
\multicolumn{5}{c} {Minimum total pressure} \\ 
\hline
Trappist-1b & 4.28$\times$10$^{3}$  & 2.40$\times$10$^{6}$   &  $\left(-803,80,10\right)$   &  $\left(-2206,96,-64\right)$ \\ 
Trappist-1c & 2.09$\times$10$^{3}$  & 2.35$\times$10$^{6}$   &  $\left(-871,68,-3\right)$   &  $\left(-1192,63,-64\right)$ \\ 
Trappist-1d & 1.06$\times$10$^{3}$  & 2.00$\times$10$^{6}$   &  $\left(-923,56,-5\right)$   &  $\left(-641,41,-59\right)$ \\
Trappist-1e & 5.68$\times$10$^{2}$  & 1.72$\times$10$^{6}$   &  $\left(-972,50,3\right)$   &  $\left(-379,32,-38\right)$ \\ 
Trappist-1f & 2.93$\times$10$^{2}$  & 1.44$\times$10$^{6}$   &  $\left(-1017,44,19\right)$   &  $\left(-218,23,-25\right)$ \\ 
Trappist-1g & 2.05$\times$10$^{2}$  & 1.30$\times$10$^{6}$   &  $\left(-1032,40,2\right)$   &  $\left(-159,20,-19\right)$ \\ 
Trappist-1h & 1.09$\times$10$^{2}$  & 1.07$\times$10$^{6}$   &  $\left(-1049,37,-16\right)$   &  $\left(-89,14,-11\right)$ \\ 
\hline
\end{tabular}
\end{table}

In Table \ref{tableSW}, note that $N_{sw}$, $T_{sw}$ and $V_{sw}$ denote the number density, temperature and velocity of the stellar wind respectively at the locations of the seven planets, whereas IMF denotes the Interplanetary Magnetic Field. The case with minimum total pressure corresponds to the fast solar wind, which features a higher velocity albeit with a lower density, whereas the converse is true for the maximum total pressure that can be associated with the slow solar wind.

\begin{figure}[!ht]
\centering
\includegraphics[scale=0.4]{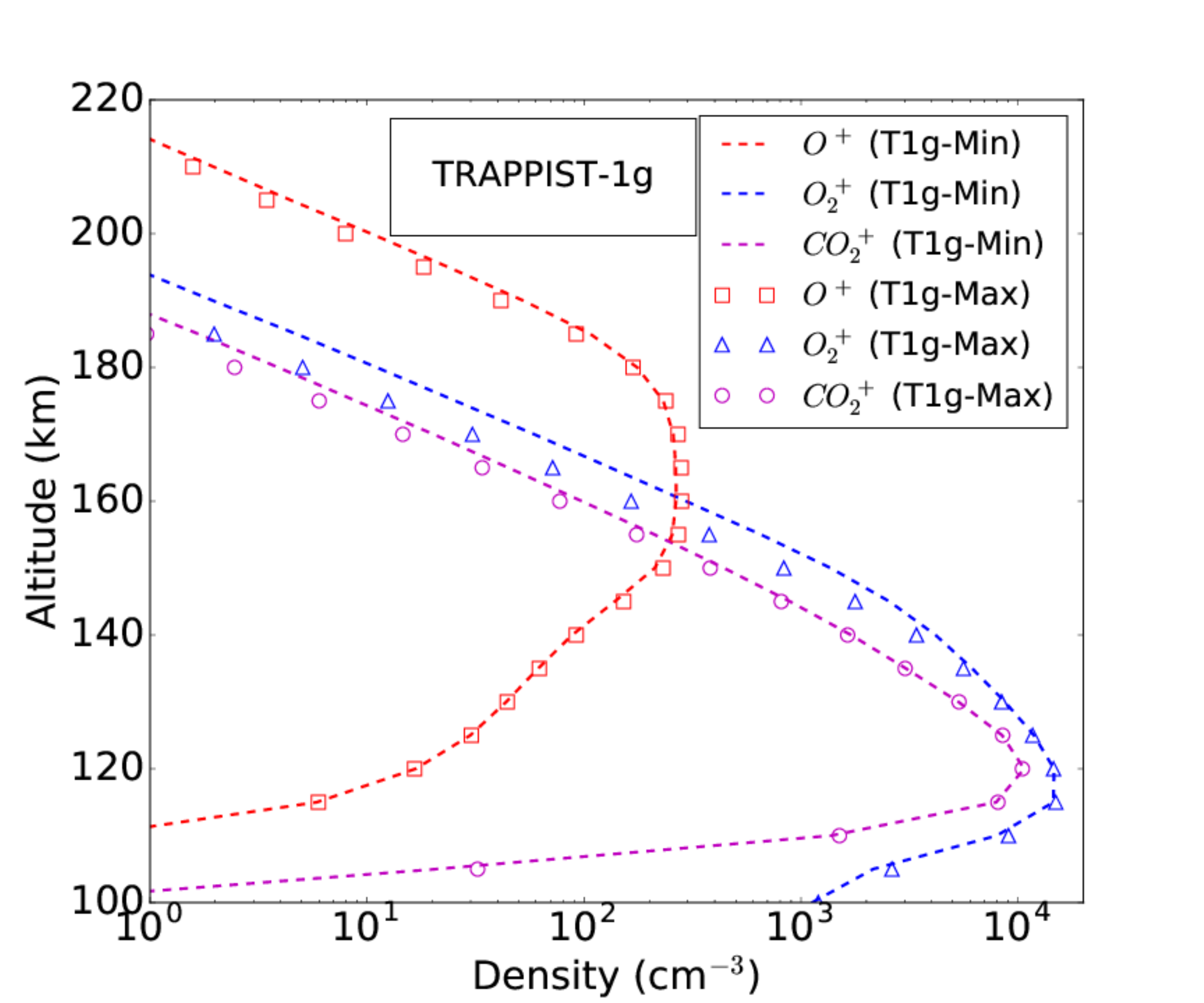} 
\caption{The ionospheric profiles along the substellar line for TRAPPIST-1g for the cases of (i) maximum and (ii) minimum total pressure over its orbit.}
\label{figIonProf}
\end{figure}

In Fig. \ref{figIonProf}, the ionospheric profiles for TRAPPIST-1g have been provided for the cases with minimum and maximum total pressure. An inspection of Table \ref{tableSW} reveals that the cases with minimum and maximum total pressure have very different stellar wind parameters, for e.g. density, velocity and interplanetary magnetic field. Despite the considerable variability in the stellar wind parameters, it is evident that the ionospheric profiles remain mostly unaffected. This would appear to indicate that the lower regions are effectively immune to the effects of the stellar wind, which lends some credibility to the fact that surface biological processes (if present) may not be significantly affected.

\end{document}